# Honesty can be the best policy within quantum mechanics

**Arindam Mitra**

Anushakti Abasan, Uttar Phalguni-7, 1/AF, Salt Lake,

Kolkata, West Bengal, 700064, India.

Abstract: Honesty has never been scientifically proved to be the best policy in any case. It is pointed out that only honest person can prevent his dishonest partner to bias the outcome of quantum coin tossing.

Honesty is said to be the best policy. However, in many cases honest persons are the loser and dishonest persons are the winner. In support of the honest losers let us imagine a case where honest person scientifically defeats a dishonest person but dishonest person can not. Then honesty can be claimed to be the best policy for that case.

Cheating and dishonesty are synonymous. In most of the cheating cases, a person is cheated by another person when they stay together at a place. To study the cheating problem, coin tossing is a good example. Suppose two parties want to take a win-loss decision by tossing a real coin. But there is no guarantee that the outcome would be cheating-free. This is true even for an ideal coin because parties may not be honest/ideal. Irrespective of the number of coins used in tossing, there is no guarantee that one would not bias the outcome. Similarly, there is always a possibility that a magician could successfully befool a spectator. In general, in any two or multi-party engagement there is always a risk of being cheated and befooled.

In 1981, Blum suggested [1] that parties should stay at different places if they wish to generate unbiased result. Essentially, he conjectured that only in on-line coin tossing parties can be mathematically compelled to become honest. This was an interesting conjecture. But it is not clear why on-line tossing might generate reliable outcome. There are many examples where a person is deceived by a distant person. So far cheating is concerned both the local and non-local environments seem to be unsafe. Be that as it may, Blum's idea is very much relevant to all two-party observations including two-party Bell's inequality test under Einstein's locality condition where parties should stay at different places. Henceforth, by coin tossing we shall mean distant coin tossing.

Suppose, Alice and Bob are separated on the question who would christen their newborn baby. To take this type of win-loss decision they would like to generate preferred outcome. If decision making is not required they may just want to generate an unbiased random outcome. For these two purposes coin tossing algorithm/protocol can be designed. A coin tossing algorithm/protocol which generates bit 0 or 1 with probability $p = \frac{1}{2} + \varepsilon$ where $\varepsilon$ is called as bias.

EPR. states always generate random data. Therefore, one can think that distant parties should share EPR states to generate uncontrollable data. However, on the basis of the assumption that shared entangled state cannot be verified [2] the possibility has been ruled out. On the basis of this assumption it has been further claimed that [2,3] zero-bias cannot be achieved within quantum mechanics. Nevertheless, in some quantum coin tossing algorithms bias around 0.2 has been achieved [4-9] and one of such algorithms has been experimentally verified [10]. Although the achieved bias is quite high, still the achievement is noteworthy since no classical algorithm with $\varepsilon$ less than 0.5 has been found.

In the impossibility proof [2] it has been rightly pointed out that without the verification of shared entanglement it cannot be used to generate unbiased result. Let us clarify this point. Suppose after preparing EPR states an honest party sends them to a dishonest party. Now sender can claim that receiver cannot bias the outcome. But the problem is, if receiver cannot verify shared entanglement he can reject sender's claim. In any two-party game, both winner and looser should know who is the winner.

There is also some lack of clarity in the assumption [2]. It is impossible to statistically verify a single shared entangled state because no-cloning principle [11,12] does not allow to produce identical copies of an unknown shared entangled state. But, statistical verification of some shared entangled states cannot be ruled out by no-cloning principle. If quantum mechanics does not allow verification of shared entanglement, then conceptual foundation of the experiments/schemes based on shared entanglement would be weak at the first place. We shall see that shared entanglement can be statistically verified.

Suppose Alice sends EPR particles to Bob. So Bob should verify shared entanglement. Otherwise Alice can cheat Bob by not sending EPR particles. But, if Bob wants to verify shared entanglement then the concern is, Bob can cheat Alice by manipulating the verification step. This is a catch -22 situation.

To evade this problem, shared entanglement is needed but it is important to *suppress* it for the time being. We shall see that by *remote* unitary operation it is possible to *lock* and *unlock* EPR correlation whenever needed. Alice can prevent Bob to bias any outcome by *remotely* locking and unlocking EPR correlation.

Let us first consider the following four EPR-Bell states of two spin-1/2 particles.

$$|\Psi_{\pm}\rangle = \frac{1}{\sqrt{2}}(|01\rangle \pm |10\rangle)$$

$$|\varphi_{\pm}\rangle = \frac{1}{\sqrt{2}}(|00\rangle \pm |11\rangle)$$

where 0 and 1 denote two opposite spin directions. Let us recall the well-known identity: $\sigma_x^2 = \sigma_y^2 = \sigma_z^2 = I^2 = I$ where $\sigma_x, \sigma_y, \sigma_z$ are three Pauli matrices and I is $2 \times 2$ identity matrix. This identity implies that qubit will remain *invariant,* if the same unitary operator $U_i \in \{\sigma_x, \sigma_y, \sigma_z, I\}$ is doubly applied on it. It is easily seen that EPR state will also remain *invariant,* if the same $U_i \in \{\sigma_x, \sigma_y, \sigma_z, I\}$ is doubly applied on one particle of an EPR pair. Evolutions of EPR singlet $|\Psi_-\rangle_i$ under the double unitary operations following the above identity may be depicted as

$$|\psi_-\rangle \xrightarrow{\sigma_x} |\varphi_-\rangle \xrightarrow{\sigma_x} |\psi_-\rangle$$

$$|\psi_-\rangle \xrightarrow{\sigma_y} |\varphi_+\rangle \xrightarrow{\sigma_y} |\psi_-\rangle$$

$$|\psi_-\rangle \xrightarrow{\sigma_z} |\psi_+\rangle \xrightarrow{\sigma_z} |\psi_-\rangle$$

$$|\psi_-\rangle \xrightarrow{I} |\psi_-\rangle \xrightarrow{I} |\psi_-\rangle$$

Next we shall present our quantum coin tossing algorithm where method of verification of shared entanglement and that of *locking* and *unlocking* EPR correlation through the above unitary operations can be realized through the algorithmic steps.

1. Alice prepares n singlets $|\Psi_-\rangle_{i=1,2,3.....n}$.
2. Alice applies unitary operator $U_i \in \{\sigma_x, \sigma_y, \sigma_z, I\}$ on one particle $A_i$ of each of her n pairs $A_iB_i$ at random and keeps the record. The resulting states are $|\varphi_-\rangle_i$, $|\varphi_+\rangle_i$, $|\Psi_+\rangle_i$ and $|\Psi_-\rangle_i$ [13]. Alice stores the particles $A_i$ in her quantum computer.
3. Alice transmits the partners $B_i$ to Bob. But Alice never discloses which $U_i$ she has applied on which particle. Bob stores the particles in his quantum computer.
4. Bob chooses $\frac{n}{2}$ particles $B_i$ at random and after disclosing the choices he requests Alice to apply the same $U_i \in \{\sigma_x, \sigma_y, \sigma_z, I\}$ on the partners $A_i$ of his chosen particles $B_i$ to convert the chosen pairs $A_iB_i$ to singlets.
5. Alice applies the same $U_i \in \{\sigma_x, \sigma_y, \sigma_z, I\}$ on the selected particles $A_i$ to convert the chosen pairs $A_iB_i$ to singlets and informs Bob when she completes the task.
6. Bob measures spin component of each of his selected particles $B_i$ along an axis $a_i$ chosen from a uniform distribution of infinite number of axes at random. Bob then requests Alice to measure spin component of each of the partners $A_i$ of his selected particles $B_i$ along his chosen axis $a_i$.
7. Alice measures spin component of each of the partners $A_i$ of Bob's selected particles $B_i$ along Bob's chosen axis $a_i$ and reveals the results to Bob.

8. Bob verifies shared entanglement with 100% correlated data. After Bob's verification of shared entanglement they proceed to generate final outcome.

9. Alice applies the same unitary operator $U \in \{\sigma_x, \sigma_y, \sigma_z, I\}$ on her remaining $\frac{n}{2}$ particles to convert the remaining $\frac{n}{2}$ shared pairs to singlets. Alice can convert the final pairs into triplets, if she wants.

10. Both measure spin component of their remaining $\frac{n}{2}$ particles along a predetermined axis, say z-axis, to generate anti-correlated data.

Let us point out the following points.

● The algorithm outputs many EPR coins at a time. However, Alice and Bob can toss a single EPR coin, if they want to generate a single bit. Of course before sharing any EPR pair they have to declare that anti-correlated data will be their final data and the bit values representing "up" and "down" spins.

● If Alice and Bob always share singlets Bob can cheat Alice by manipulating the verification step. Bob can secretly measure spin component of each of his particles along z-axis. In step 6, Bob requests Alice to measure spin component of her particles whose partners already gave him 0s. In step 7, if Alice performs measurement on Bob's chosen particles along z-axis she will get 1s. So Alice's final particles will yield 0s whose partners gave Bob 1s. It means Bob can choose 0 or 1 as he wants.

Suppose Alice is honest and Bob is dishonest. In this condition, bias $\varepsilon_B$ given by Bob can be made 0, no matter whether Bob measures or acts as it is described in the above steps. Even Alice can allow Bob to choose any pair from the final set to generate a bit. Next we shall see that Alice can completely *lock* EPR correlation so that Bob cannot know or control Alice's result 0 or 1 with probability more than 1/2.

***Proof:*** The density matrix of their shared EPR states prior to the disclosure of Bob's choice of particles (upto step 4) may be written as

$$\rho_{AB} = \frac{1}{4}\left(|\psi_-\rangle\langle\psi_-| + |\psi_+\rangle\langle\psi_+| + |\varphi_-\rangle\langle\varphi_-| + |\varphi_+\rangle\langle\varphi_+|\right)$$

$$= \tfrac{1}{8}[(|01\rangle - |10\rangle)(\langle 01| - \langle 10|) + (|01\rangle + |10\rangle)(\langle 01| + \langle 10|) + (|00\rangle - |11\rangle)(\langle 00| - \langle 11|) + (|00\rangle + |11\rangle)(\langle 00| + \langle 11|)]$$

$$= \frac{1}{4}\left(|01\rangle\langle 01| + |10\rangle\langle 10| + |00\rangle\langle 00| + |11\rangle\langle 11|\right)$$

$$= \frac{1}{4}I,$$ where I is $4\times 4$ identity matrix. Note that $\rho_{AB}$ also represents the equal *mixture* of four direct product states $|01\rangle, |10\rangle, |00\rangle,$ and $|11\rangle$ (here $|01\rangle = |0\rangle_A|1\rangle_B$, and so on.).

Alice never reveals which unitary operator $U_i$ was applied on which singlet to prepare $\rho_{AB}$. Therefore, prior to the disclosure of his choice of particles (upto step 4) it is impossible for Bob to know whether $\rho_{AB}$ is a *mixture* of the above entangled states or direct product states. Equally probable direct product states $|0\rangle_A|1\rangle_B, |1\rangle_A|0\rangle_B, |0\rangle_A|0\rangle_B$ and $|1\rangle_A|1\rangle_B$ always generate totally uncorrelated data. It implies that upto the step 4 Bob's probability of controlling Alice's result is 1/2.

Alice *locks* entanglement by preparing $\rho_{AB} = \frac{1}{4}I$. After the step 4 Alice *unlocks* EPR correlation by converting $\rho_{AB}$ into singlets/triplets. Of course this conversion is possible, if Bob does not secretly measure or apply $U_i$ at random on all his particles upto the step 4. Whatever Bob does, after the step 4 Bob's probability of controlling Alice's result is 1/2. Therefore, Bob's probability of generating a preferred bit is always $p_B = \frac{1}{2}$. It means $\varepsilon_B = 0$. As Alice cannot know which particle Bob will choose she is bound to use singlets if she is assumed to be honest. That is, $\varepsilon_A = 0$ if Alice is assumed to be honest.

Suppose Alice is dishonest. Therefore, Alice is not bound to use 100 % singlets. Suppose in the first step Alice uses 50% known unentangled pure direct product states $|1\rangle_i |1\rangle_i$ and 50% singlets. Up to the step 4, their shared state can be described as $\rho^*_{AB} = \frac{1}{2}|1\rangle\langle 1| + \frac{1}{2}\rho_{AB}$ where $\rho_{AB} = \frac{1}{4}I$. On this state

Alice as well as Bob's probability of generating bit 1 is $p = \left(\frac{1}{2} + \frac{1}{2} \cdot \frac{1}{2}\right)$. It means for bit 1, $\varepsilon_A = \varepsilon_B = 1/4$. It can now be concluded that only honest Alice can prevent dishonest Bob to bias the outcome. That is, only honesty can defeat dishonesty within quantum mechanics. Here dishonesty is asymmetrically prohibited within quantum mechanics.

In this algorithm, honest sender achieves zero-bias, but not the honest receiver. Suppose Bob, the receiver, is honest. For the above mentioned example, honest Bob is compelled to bias the outcome with probability 1/4 if Alice is dishonest and on the other hand dishonest Alice can bias the outcome with probability 1/4 if Bob is honest. Neither Bob can prevent Alice to bias the outcome nor he can prevent himself from giving bias.

Let us point out the following points.

● It can be pointed out that n need not to be arbitrarily large number. Here, n needs to be greater than one for the verification purpose. But shared entanglement can be verified with minimal statistics.

● Here Alice, the sender, is getting advantage. If Alice and Bob interchange their role then honest Bob could prevent dishonest Alice to bias the outcome.

● The algorithm will crash if both parties stay at the same place. The reason is simple. Bob cannot reliably verify shared entanglement in presence of Alice. On the other hand Alice cannot reliably suppress entanglement in presence of Bob. So, the final pairs cannot be considered as EPR pairs if they stay together. One can cheat on other's observation. Due to this reason, cheating-free outcome cannot be generated locally. The nature of the problem demands an on-line solution of the problem as Blum suggested [1]. But without verification nothing can be trusted. Shared entanglement can be verified at a distance.

●In step 1, Alice prepares pure singlets. Needless to say, quantum mechanics does not forbid preparation of pure state. But, if Alice collects some impure singlets, still she can produce pure singlets from the impure ones by entanglement purification technique [14]. As noise is always present, singlets have to be considered as impure singlets when they will share them after step 4. The proof demands that the final shared state has to be pure singlet. To produce pure singlets from impure ones parties have to again purify them. But the problem is, no cheating-free two-party entanglement purification algorithm exists.

In conclusion, the remaining task is to achieve zero-bias symmetrically in ideal as well as non-ideal condition. On the basis of the presented algorithm and alternative quantum encoding [15,18], the task can be accomplished [19] where both parties will be compelled to be honest.

**Note added:** On an earlier version one of the referees observed that paper "*deserve to be widely read and analyzed*". Interested readers may see my other works [20-21].

*Email: mitra1in@yahoo.com